\documentstyle[12pt]{article}
\begin{document}
\vspace*{2.0in}
\begin{center}
{\LARGE Intermittency in soft hadronic processes and Zip--model}
\end{center}
\vspace{0.2in}
\centerline{\bf  E.~G.~Gurvich\footnote{E--mail address:
gurvich@taunivm}, G.~G.~Leptoukh\footnote{E--mail address:
visitor@pyrssc.physics.ncsu.edu}, E.~K.~Sarkisyan\footnote{E--mail address:
saek@physics.aod.ge},}
\begin{center}
{\it Institute of Physics, 6 Tamarashvili st., Tbilisi--77,
Republic of Georgia}
\end{center}

\begin{abstract}

A low constituent number scheme based on the nontrivial gluon string
splitting (the Zip--model) is shown to yield a substantial intermittency
for soft hadronic processes. With a simplest addition of the
Bose--Einstein correlations the remarkable agreement with the NA22
experimental data on rapidity factorial moments is reached.

\end{abstract}
\pagestyle{empty}
\newpage
\pagestyle{plain}
\setcounter{page}{1}

\section{Introduction}

Recently experimentally observed intermittent structure of
produced--particle distributions in high energy collisions
\cite{dremin} and frustrating situation in the theoretical
description of the phenomena \cite{dremin,pesch} makes it crucial in
the choice of the multiple production process model.

In this note we study the problem for soft hadronic interaction within
 the low constituent number model \cite{agl} with the nontrivial gluon
string hadronization (the Zip--model) \cite{gl}. Such an approach
allows one to obtain reasonable multiple hadron production properties
in addition to a successful description \cite{land} of the total and
elastic cross--sections within a single pomeron exchange picture.

The reason we expect to obtain intermittent patterns in this
model is two--fold.
First, the standard quark string hadronization stage seems to be a
real intermittency source phase \cite{dremin}, and the model, below
considered, mostly deals with this regime.
Second, additional multi--particle short--range correlations in our
model due to the Zip--mechanism lead to an enhancement of intermittent
behavior of particle distribution.

Following ref.\cite{bp} intermittency refers to a scaling law,

$$
F_i\, \propto M^{\varphi_i},
\qquad 0<\varphi_i\leq i-1,
$$
of the scaled factorial moments,

$$
F_i =
\frac{1}{M}
\sum^{}_{cells} \frac{
\left
\langle n\, (n-1)\cdots (n-i+1)
\right \rangle
}
{{\langle n \rangle}^i}
$$
as a function of $M$ cells of particle phase space. $\varphi_i$, called
the intermittency index, shows a strength of intermittent behavior.
Here the brackets denote averaging over events and $n$ is
multiplicity in the chosen cell.

We are to show here that contrary to the popular two--chain models (FRITIOF
\cite{fr} and two--chain DPM \cite{dpm}) our Zip--model due to intrinsic
positive multiple correlations leads to intermittent behavior
being compatible with the NA22 experimental observations \cite{na22}
after count of Bose--Einstein interference (Sec.3).
The brief description of the Zip--model is given in Sec.2.

\section{Zip--model of soft hadronic interaction}

In the Zip--model \cite{agl,gl} we consider a soft hadronic interaction
picture consisting of several stages. A hadron wave function is created
far from the interaction point. The number of projectile constituents
taking part in a soft hadron interaction is restricted by the valence
component only. The constituent momenta are not changed during the
interaction. An interaction goes via color exchanges, and despite
their dynamical properties the colored objects rush out of the
interaction region and stretch strings.  The resulting strings can be
divided into classes by their end--point color "charges" ($SU(3)_c$
representations). The gluon  (adjoint) string dominates in the final
state. The fragmentation of a gluon string differs from that of two
independent quark strings causing the required properties of
soft processes.

The gluon string splits into two quark ones by the certain
Zip--mechanism --- locally via creation of the so--called zip--antizip
pairs. Each zip is a certain junction between a gluon and two quark
strings, switching strings from the adjoint representation of $SU(3)_c$
group into the fundamental one and vice versa.  These massless
objects are created uniformly along the gluon string with a certain
$\omega$--probability (just the same way as $q\bar q$ are produced
along the usual quark string with $w$--probability \cite{artru,cnn,egg}).

The very moment a $z \bar z$--pair is produced, two very short quark
strings appear out of the gluon string. They can be naturally treated
as ordinary ones with ends stretched by zip and antizip instead of
quark and antiquark.  We suppose that zip and antizip from the adjacent
$z \bar z$ annihilate. Thus after this "unzipping" process, we
obtain finally almost the same two long quark strings as in the standard
approach.

The created quark strings break down via the usual $q \bar q$ creation
\cite{artru,lund}. Quarks and antiquarks from adjacent $q \bar
q$--pairs meet each other and form the so--called primary fragments.
What is important here is that $z \bar z$--pair and $q \bar q$--pair
creation processes overlap in the space--time leading to substantially
positive correlations that are seen in the experiment \cite{agl}.
Certainly, such a collective effect can lead to quite an intermittent
behavior for soft hadronic process.

The above--mentioned scheme has been described in ref.\cite{gl} at
the level of primary fragments, while the main details of the
Monte--Carlo version are outlined in ref.\cite{agl}. We've used the same
string breaking algorithm both for the "zip" part and for a quark
string one. To go from primary fragments to real hadrons we've used
an available HERWIG 4.6 \cite{her} cluster--hadron converter without
any modifications besides the processing of high mass cluster that we
suppose to decay as ordinary quark string.

\section{Results and Discussion}

In Fig.1 we present results for various order scaled factorial moments
in rapidity space obtained by our MC--generator for $\omega / w = 0.1$
(dashed line) as well as for $\omega=\infty$ (dotted line).  The latter
corresponds to the case of two independent quark strings, i.e.  to the
DPM--like two--chain model.  We see that in fact our model develops
higher magnitudes for all the moments considered than the usual
two--chain model.  Note also that switching on of the Zip--mechanism
gives quite a visible positive change for high order factorial moments
especially and, what is important, for small rapidity bins as can be
seen in Fig.1d.

The reason for such an increase of factorial moment magnitude in our model
is connected to the fact that short quark strings stretched by $z \bar z$
are created in pairs, so particles out of these quark string pairs are to be
correlated.  This indicates an existence of genuine multi--particle
fluctuations at a gluon string hadronization stage.

Nevertheless, it is seen that the Zip--model by itself does not describe
the experiment (error bars in Fig.1) \cite{na22} despite the evident
excess over the independent two--chain version.

Meanwhile, it has been shown recently \cite{Kit} that Bose--Einstein
(BE) interference plays an important role in the intermittency effect
in addition to a mechanism causing the underlying power--law behavior.

Indeed, when taking into account BE interference in the simplest
possible way (using the corresponding LUND routine \cite{jetset} with default
parameters) we obtain the remarkable agreement with the data for
various order scaled factorial moments, as shown in Fig.1 by solid
lines.

The corresponding intermittency indices computed in our model for
various values of gluon string splitting relative parameter $\omega
/w$ are listed in Tab.1.  Although the low order moment behavior
seems to be almost the same for different $\omega /w$, the high order
moments (especially, $F_5$) are consistent with experiment for the
certain $\omega /w = 0.1$ contrary to the two independent quark string
case.

It is worth noticing that the pure BE--effect by itself is insufficient
to agree with the experimental magnitudes of the moments
considered.

{}From the obtained results we conclude that our Zip--approach combined
with Bose--Einstein interference is able to fit the observed
intermittent behavior in soft hadronic processes.

\section{Acknowledgments}
Authors are indebted to I.~Dremin, W.~Kittel and Yu.~Werbetsky for
valuable discussions and M.~Charlet for providing us with the
required NA22 data and processing routines.

\newpage
\begin{flushleft}
{\bf Table 1}
\end{flushleft}
\smallskip
Intermittency indices $\varphi_i$ from NA22 experiment \cite{na22}
and Zip--model (BE interference included) results at different
parameter $\omega /w$ values.

\begin{flushleft}
\begin{tabular}{|c|c|c|c|c|c|c|} \hline  \hline
$i$&$\omega /w=\infty$
&$\omega /w=1$ &$\omega /w=0.5$ &$\omega /w=0.1$& experiment\\ \hline
 2&$ 0.022\pm  0.002$&$   0.019\pm  0.002$&$  0.016\pm  0.002$&$
      0.014\pm  0.002$&$0.018\pm 0.001$\\
 3&$ 0.062\pm 0.007$&$    0.056\pm  0.004$&$  0.061\pm  0.004$&$
      0.055\pm  0.005$&$0.055\pm 0.002$\\
 4&$ 0.153\pm  0.017$&$   0.123\pm 0.011$&$   0.120\pm  0.010$&$
     0.153\pm   0.011$&$0.14 \pm 0.01$ \\
 5&$ 0.150\pm  0.042$&$   0.175\pm 0.028$&$   0.134\pm  0.023$&$
0.271\pm   0.026$&$0.30 \pm 0.02$\\ \hline
\end{tabular}
\end{flushleft}
\newpage
\thispagestyle{empty}
{\Large \bf Figure captions}\\
\\
{\bf Fig.1} \  Scaled factorial moments in rapidity space.
The data are from NA22 experiment \cite{na22}, while lines represent
Zip--model results for various values of $\omega /w$.

\end{document}